\documentclass[a4paper,twocolumn]{quantumarticle}
\pdfoutput=1
\usepackage[utf8]{inputenc}
\usepackage{amsmath,amsxtra,amssymb,amsthm,amsfonts}
\usepackage{mathrsfs}
\usepackage[T1]{fontenc}
\usepackage{mathtools}   
\usepackage{multirow}
\usepackage{tikz,quantikz}
\usepackage{physics}
\usepackage{bbm}
\usepackage{subfigure}
\usepackage{comment}
\usepackage{appendix}
\usetikzlibrary{arrows, automata}
\usepackage{color,hyperref}
\definecolor{dblue}{rgb}{0, 0, 0.72}

\allowdisplaybreaks
\numberwithin{equation}{section}
\newtheorem{lemma}{Lemma}[section]
\newtheorem{prop}[lemma]{Proposition}
\newtheorem{theorem}[lemma]{Theorem}

\newtheorem{rem}[lemma]{Remark}

\newtheorem{example}[lemma]{Example}
\newtheorem{definition}[lemma]{Definition}

\newcommand{\re}{\begin{rem}\rm}
\newcommand{\mar}{\end{rem}}

\newcommand{\qd}{\end{proof}\vspace{0.5ex}}
\newcommand{\mb}{\mathbb}
\newcommand{\mc}{\mathcal}
\newcommand{\adm}{\text{ad}}
\newcommand{\id}{\text{id}}

\newcommand{\added}[1]{#1}
\begin{document}
\title{Resource-Dependent Complexity of Quantum Channels}
\author{
  Roy Araiza$^{1}$,
  Yidong Chen$^{2}$,
  Marius Junge$^{1}$,
  Peixue Wu$^{3,*}$
}

\affiliation{
  $^{1}$Department of Mathematics and Illinois Quantum Information Science and Technology Center (IQUIST), University of Illinois at Urbana-Champaign \\
  $^{2}$Department of Physics, University of Illinois at Urbana-Champaign \\
  $^{3}$Institute for Quantum Computing and Department of Applied Mathematics, University of Waterloo
}

\thanks{R.A. was supported as a J.L. Doob Research Assistant Professor.}
\thanks{M.J. was partially supported by NSF Grants DMS-2247114, DMS-1800872, and NSF RAISE-TAQS-1839177.}
\thanks{P.W. is the corresponding author and was supported by the Canada First Research Excellence Fund (CFREF)}

\begin{abstract}
We introduce a new framework for quantifying the complexity of quantum channels, grounded in a suitably chosen resource set. This class of convex functions is designed to analyze the complexity of both open and closed quantum systems. By leveraging Lipschitz norms inspired by quantum optimal transport theory, we rigorously establish the fundamental properties of this complexity measure. The flexibility in selecting the resource set allows us to derive effective lower bounds for gate complexities and simulation costs of both Hamiltonian simulations and dynamics of open quantum systems. Additionally, we demonstrate that this complexity measure exhibits linear growth for random quantum circuits and finite-dimensional quantum simulations, up to the Brown-Susskind threshold.
\end{abstract}
\maketitle
\section{Introduction}\label{section:intro}
\noindent \textit{Background.}---Determining the resources required to perform a quantum task is a fundamental question in quantum information theory. In this letter, we aim to provide certificates for the complexity of quantum operations.

Certifying that a large number of elementary gates is required to approximate a given unitary operation is a challenging task in the study of gate complexity. This difficulty is compounded by the existence of numerous, often nonequivalent, notions of quantum complexity measures \cite{OT, RW, JW, FR, PD, KH, BT, BV, BASST, BGS, BSh}. Beyond gate counts, these complexity measures have been extended to broader contexts, including quantum many-body systems \cite{JM, HR, BS1, BS2, MA} and chaotic quantum systems \cite{PCASA, DS1, ADS, RSSS1, RSSS2}. A pivotal development in this field was made by Brown and Susskind in \cite{BS1, BS2}, where they conjectured that the complexity of a typical random quantum circuit grows linearly with its depth before saturating at a value exponential in the system size. This conjecture underscores the intrinsic incompressibility of quantum circuits and the fundamental difficulty of performing most quantum operations within finite resources.

Motivated by this conjecture, researchers have focused on establishing sharper lower bounds for gate complexities. Notable progress has been made in proving the \textit{Brown-Susskind} conjecture in specific settings, such as gate counts for different designs of random quantum circuits \cite{HFKEH, Haferkamp23, chen2024incompressibility}. Furthermore, in \cite{Halpern22}, the conjecture was validated from a resource-theoretic perspective, by defining and analyzing a resource theory for uncomplexity. These advancements have significantly enriched our understanding of quantum complexity and its implications for both theoretical and practical aspects of quantum computation.

\noindent \textit{Our contribution.}---We introduce a flexible, resource-dependent quantum complexity measure inspired by the axiomatic approach to complexity considered in \cite{LBKJL} and ideas from quantum optimal transport theory \cite{depalma2023quantumoptimaltransportquantum, CM1, CM2, Wir2, Zyczkowski_1998, caglioti2021optimaltransportquantumdensities, gaorouze24} and noncommutative geometry \cite{AC, rieffel1999metrics, rieffel2004compact}. The novelty of our approach lies in the flexibility of choosing resource sets and the ability to treat both closed and open quantum systems. This flexibility allows us to recover several existing complexity measures, including the Wasserstein complexity introduced in \cite{LBKJL} and the geometric complexity introduced in \cite{N1, N2, N3}.

Our complexity measure possesses desirable mathematical properties such as convexity, additivity under tensor products, and Lipschitz continuity. Notably, it provides a lower bound for the gate count, enabling us to propose effective lower bounds on the number of gates required to accomplish a given task. Our method is completely different from the techniques developed in \cite{hahn2024lowerboundstrottererror, rosenthal2024quantumchanneltestingaveragecase}. 

Finally, we prove a version of the Brown-Susskind conjecture, demonstrating the linear growth of complexity for a random unitary drawn from the unitary resource set $S$. We also establish a similar lower bound on the complexity of Hamiltonian simulation via the quantum stochastic drift (qDRIFT) protocol considered in \cite{Campbell19}, as implemented in Qiskit \cite{Qiskit}.


\noindent \textit{Definition of the Complexity Measure}.---Suppose $\mathcal{H}$ is a finite-dimensional quantum system (Hilbert space). Denote:\\
- $\mathcal{D}(\mathcal{H})$ as the set of density operators on $\mathcal{H}$, corresponding to quantum states.\\
- $\mathcal{B}(\mathcal{H})$ as the set of linear operators on $\mathcal{H}$.\\
- $\mathcal{O}(\mathcal{H})$ as the set of Hermitian operators, corresponding to observables in quantum mechanics.

We say that $S$ is a \emph{resource set} on system $\mathcal{H}$ if $S$ is a subset of $\mathcal{B}(\mathcal{H})$. For any resource set $S$, we can define an induced \textit{Lipschitz semi-norm} on $\mathcal{B}(\mathcal{H})$ given by
\begin{equation}\label{eqn:Lipschitz_norm}
    |||x|||_S := \sup_{s \in S} \|[s, x]\|_{\mathrm{op}}, \quad x \in \mathcal{B}(\mathcal{H}),
\end{equation}
where $[s, x] = sx - xs$ is the commutator, and $\|\cdot\|_{\mathrm{op}}$ is the operator norm, given by the largest singular value of the operator. \added{This semi-norm is motivated from noncommutative geometry, initiated by Connes \cite{connes1992metric}.}

When $S$ contains only Hermitian operators, we can consider $s \in S$ as generating a unitary evolution $x(t) = e^{i t s} x e^{-i t s}$ on observables (Heisenberg picture). The commutator relates to the time derivative of the evolution of observables at time $t = 0$:
\begin{equation}
    \left.\frac{d}{d t} x(t) \right|_{t = 0} = i [s, x].
\end{equation}
Therefore, $|||x|||_S$ represents the maximal rate at which the observable $x$ changes under the dynamics generated by the elements of $S$.

\added{The Lipschitz semi-norm \eqref{eqn:Lipschitz_norm} naturally induces an Earth Mover's distance, also referred to as the Wasserstein-1 distance \cite{PMTL} in the context of quantum states. More generally, any semi-norm on observables induces a true distance on quantum states via the dual formulation \cite{gaorouze24, rieffel1999metrics, rieffel2004compact}. We will see a variant of \eqref{eqn:Lipschitz_norm} in Section \ref{sec:geometric}.} For any quantum states $\rho, \sigma \in \mathcal{D}(\mathcal{H})$, the Earth Mover's distance is defined by
\begin{equation}
    W_S(\rho, \sigma) = \sup \left\{ \left| \operatorname{tr}(x (\rho - \sigma)) \right| : |||x|||_S \leq 1, \ x = x^\dagger \right\}.
\end{equation}
In an $n$-qudit system, if we choose $S$ to be the set of single-qudit Pauli gates, our Earth Mover's distance $W_S(\rho, \sigma)$ recovers the Wasserstein-1 distance introduced in \cite{PMTL}. Compared to the expression of the Wasserstein-1 distance, the Lipschitz norm \eqref{eqn:Lipschitz_norm} as the dual picture is easier to compute or estimate in some cases.

We introduce the complexity measure based on the above Lipschitz semi-norm and its induced distance on quantum states. Suppose $\Phi$ is a quantum channel acting on density operators on $\mathcal{H}$, and denote $\Phi^*$ as the dual map acting on observables on $\mathcal{H}$. Denote $\operatorname{tr}$ as the trace; we have
\begin{equation}
    \operatorname{tr}(\Phi(\rho) X) = \operatorname{tr}(\rho \, \Phi^*(X)),
\end{equation}
where $\rho$ is a density operator and $X \in \mathcal{O}(\mathcal{H})$ is an observable. Given a resource set $S$, our resource-dependent complexity $C_S$ is defined by
\begin{align}
    C_S(\Phi) &= \sup_{\rho \in \mathcal{D}(\mathcal{H})} W_S\left( \Phi(\rho), \rho \right) \\
    &= \sup_{|||x|||_S \leq 1, \ x = x^\dagger} \| \Phi^*(x) - x \|_{\mathrm{op}}. \label{def:complexity_hermitian}
\end{align}

\added{From a physical perspective, $C_S(\Phi)$ captures the largest possible deviation that the channel $\Phi$ induces relative to the Lipschitz geometry defined by the resources in $S$.  \\
\indent In the Schr\"odinger picture, this means asking: for the worst-case input state, how far does $\Phi$ push it away, as measured by the Wasserstein distance associated with $S$. Equivalently, in the Heisenberg picture, $C_S(\Phi)$ quantifies the maximal disturbance of observables caused by $\Phi^*$. If an observable $x$ is simple with respect to $S$ (i.e., it has Lipschitz norm bounded by one), then $\Phi^*(x)-x$ measures how much the dynamics reshapes that observable. \\
\indent This dual interpretation highlights the operational meaning: $C_S$ is the worst-case cost to simulate or reverse $\Phi$ under the constraints of $S$. 
}

\section{Preliminaries}
\subsection{Definitions and Basics}
A quantum channel $\Phi: \mathcal{D}(\mathcal{H}) \to \mathcal{D}(\mathcal{H})$ is a linear map that is completely positive and trace-preserving (CPTP). The dual map $\Phi^*$, is unital and completely positive (UCP).

Given a resource set $S \subseteq \mathcal{B}(\mathcal{H})$, recall that we define the induced Lipschitz semi-norm \eqref{eqn:Lipschitz_norm} on $\mathcal{B}(\mathcal{H})$ as
   $ ||| x |||_S := \sup_{s \in S} \| [s, x] \|_{\operatorname{op}},\quad x \in \mathcal{B}(\mathcal{H})$, where $[s, x] = sx - xs$ denotes the commutator, and $\| \cdot \|_{\operatorname{op}}$ is the operator norm.

The environment-assisted (also known as complete in \cite{Paulsen2003}) Lipschitz norm is defined when we allow the quantum system to be coupled with an additional $d$-dimensional quantum system $\mathcal{H}_d$:
\begin{equation}
    ||| X_d |||_S^d := \sup_{s \in S} \| [s \otimes \mathbb{I}_d, X_d] \|_{\operatorname{op}},\quad X_d \in \mc B(\mathcal{H} \otimes \mathcal{H}_d),
\end{equation}
where $\mathbb{I}_d$ denotes the identity operator on $\mathcal{H}_d$. 

Then, the resource-dependent complexity of a quantum channel $\Phi$ is defined as $C_S(\Phi) = \sup_{||| x |||_S \le 1,\ x = x^{\dagger}} \| \Phi^*(x) - x \|_{\operatorname{op}}.$ For notational simplicity, we will omit the condition $x = x^{\dagger}$ from now on.

We also introduce the environment-assisted resource-dependent complexity, which has better mathematical properties and can provide us with sharper lower bounds in some applications. It is defined by
\begin{equation}
    C_S^{\operatorname{cb}}(\Phi) = \sup_{d \geq 1} \sup_{||| X_d |||_S^d \leq 1} \| (\Phi^* \otimes \operatorname{id}_{\mc B(\mathcal{H}_d)} )(X_d) - X_d \|_{\operatorname{op}},
\end{equation}
where $\operatorname{id}_{\mc B(\mathcal{H}_d)}$ is the identity super-operator on $\mc B(\mathcal{H}_d)$. As a remark, $cb$ is the abbreviation of completely bounded \cite{Paulsen2003}. 

To simplify notation, following the convention in \cite{Paulsen2003}, the above can be abbreviated as
\begin{equation}
    C_S^{\operatorname{cb}}(\Phi) = \sup_{||| X |||^{\operatorname{cb}}_S \leq 1} \| \Phi^*(X) - X \|_{\operatorname{cb}}.
\end{equation}

Using elementary properties such as submultiplicativity and the triangle inequality for (semi-)norms, we can show that $C_S^{\operatorname{cb}}(\Phi)$ has the following desirable properties (Since the proof is straightforward, we refer the interested reader to \cite{araiza2025} for the full mathematical details in the von Neumann algebra setting):

\begin{lemma} \label{lemma:elementary property}
\begin{enumerate}
    \item $C_S^{\operatorname{cb}}(\Phi) = 0$ if and only if $\Phi = \operatorname{id}$.
    \item \textbf{(Subadditivity under concatenation)} For any quantum channels $\Phi$ and $\Psi$, we have
    \[
        C_S^{\operatorname{cb}}(\Phi \circ \Psi) \leq C_S^{\operatorname{cb}}(\Phi) + C_S^{\operatorname{cb}}(\Psi).
    \]
    \item \textbf{(Convexity)} For any probability distribution $\{p_i\}_{i \in I}$ and quantum channels $\{\Phi_i\}_{i \in I}$, we have
    \begin{equation}
        C_S^{\operatorname{cb}}\left( \sum_{i \in I} p_i \Phi_i \right) \leq \sum_{i \in I} p_i C_S^{\operatorname{cb}}(\Phi_i).
    \end{equation}
    \item \label{universal upper bound} \textbf{(Maximal complexity given by expected length of $S$)} Under the assumption that the commutant $S'$ is a $*$-algebra, $\Phi^*|_{S'} = \operatorname{id}$, and $\kappa(S)$ is defined in \eqref{def:expected length}, we have
    \begin{equation}
        C_S^{\operatorname{cb}}(\Phi) \leq \kappa(S) \| \Phi - \operatorname{id}\|_{\diamond}.
    \end{equation}
    \item \textbf{(Tensor additivity)} For any quantum channels $\Phi$ and $\Psi$, we have
    \begin{equation}
        C_{S \vee S}^{\operatorname{cb}} (\Phi \otimes \Psi) = C_S^{\operatorname{cb}}(\Phi) + C_S^{\operatorname{cb}}(\Psi),
    \end{equation}
    where $S \vee S := \{ \mathbb{I} \otimes s_1,\ s_2 \otimes \mathbb{I} \mid s_1, s_2 \in S \}$.
\end{enumerate}
\end{lemma}

\subsection{Quantum Expected Length and Estimates of Complexity}\label{Calculation}
We first introduce the definition of the expected length $\kappa(S)$ of the resource set $S$, which provides a universal upper bound for the resource-dependent complexity of any quantum channel. Throughout this letter, we assume that 
\begin{align*}
    S' := \{ x \in \mathcal{B}(\mathcal{H}) : x s = s x,\ \forall s \in S \}
\end{align*}
is a $*$-algebra.

\added{The key technical ingredient is the \emph{conditional expectation} $E_{\text{fix}}: \mathcal{B}(\mathcal{H}) \to S'$. Intuitively, this map ``forgets'' all the information in an operator $X$ that is invisible to the algebra $S'$. More precisely, $E_{\text{fix}}$ is unital and satisfies
\begin{equation}
   E_{\text{fix}}(s_1 X s_2) = s_1 E_{\text{fix}}(X) s_2, \quad \forall s_1, s_2 \in S',\ X \in \mathcal{B}(\mathcal{H}).
\end{equation}
It is well known that such maps are unital completely positive, and hence $E_{\text{fix}}^*$ is a valid quantum channel. Appendix~\ref{app:conditional expectation} summarizes the minimal properties we need here, and we refer the reader to \cite{stormer} for a comprehensive overview.}
\added{
\begin{example}\label{ex:Pauli resource}
If $S$ is given by all single-qubit Pauli operators in the $n$-qubit system, then $S'$ consists only of multiples of the identity. In this case $E_{\text{fix}}(X) = \operatorname{tr}(\sigma X)\,\mathbb{I}$ for some state $\sigma$, so the dual channel always outputs $\sigma$, i.e.\ $E_{\text{fix}}^*(\rho) = \sigma$ for every input state $\rho$. In this case, the complexity of this conditional expectation measures the worst-case cost to map any input state to the target state $\sigma$ under the constraints of single Pauli operators. 
\end{example} }
We may now define the \textit{quantum expected length} of the resource set $S$ by
\begin{equation}\label{def:expected length}
    \kappa(S) := C_S^{\operatorname{cb}}(E_{\text{fix}}^*).
\end{equation}
\added{Physically, $\kappa(S)$ quantifies the maximum ``stretch'' needed to recover the action that erases everything except what is fixed by $S$. It is central because it sets a universal upper bound for the resource-dependent complexity of any quantum channel.}

Now we discuss how to compute or estimate the complexity measure we propose. In general, it is challenging to compute the complexity numerically (this worst-case complexity is expected to be NP-hard), but we can provide lower bound estimates by constructing a certificate observable that nearly achieves the supremum in \eqref{def:complexity_hermitian}. We illustrate a simple case introduced in Example \ref{ex:Pauli resource}: in the $n$-qubit system, suppose
\begin{align*}
S &= \{ \sigma_x^j : 1 \leq j \leq n \} \cup \{ \sigma_y^j : 1 \leq j \leq n \},
\end{align*}
where $\sigma_x^j$ and $\sigma_y^j$ are Pauli-X and Pauli-Y operators acting on the $j$-th qubit.

Using the observable $X = \sum_{j=1}^n \sigma_x^j$, we can achieve a lower bound of $\kappa(S)$ for the trace-preserving conditional expectation $E_{\text{fix}}$. Indeed, one can easily verify that $S' = \mathbb{C} \cdot \mathbb{I}_{2^n}$ and $E_{\text{fix}}(X) = \frac{\tr(X)}{2^n} \mathbb{I}$. Then,
\begin{align*}
& ||| X |||_S = \sup_{s \in S} \| [s, X] \|_{\infty} \\
& 
= \sup_{1 \leq k \leq n} \| [\sigma_y^k, \sum_{j=1}^n \sigma_x^j] \|_{\infty} = \| [\sigma_y, \sigma_x] \| = 2; \\
& \| X \|_{\operatorname{op}} = n,
\end{align*}
which provides a lower bound
\begin{align*}
    C_S(E_{\text{fix}}^*) \geq \frac{ \| E_{\text{fix}}(X) - X \|_{\operatorname{op}} }{ ||| X |||_S } = \frac{n}{2}.
\end{align*}

On the other hand, using the tensor product structure $E_{\text{fix}} = E_{\tau}^{\otimes n}$, where $E_{\tau} = \frac{\tr(X)}{2} \mathbb{I}_2$ acts on a single-qubit system, we can show via tensor additivity that $C_S^{\operatorname{cb}}(E_{\text{fix}}^*) \leq n$. In summary, the quantum expected length with a resource set given by single Pauli gates has order $n$.

\added{
As another example, we can also construct resource set 
\begin{equation}
    S = \{\ketbra{\psi_k}{\psi_k}: 1 \le k \le 2^n\},
\end{equation}
where for each $k = 1,2,\cdots, 2^n$, we define 
\begin{equation}
    \ket{\psi_k} = \ket{i_1\cdots i_n},\quad k -1 = \sum_{t = 1}^n i_t 2^{t-1}.
\end{equation}
Using the similar argument as before, we can show that $C_S(E_{\rm fix}^*) \ge c 2^n$ for some universal constant $c > 0$ using the testing operator $X = \sum_{k=1}^{2^n} k \ketbra{\psi_k}{\psi_k}$, which is a sharp contrast to Wasserstein complexity, which can achieve at most $n$ (see \cite{LBKJL}).
}

\section{Comparison of Complexities}

\subsection{Gate Complexity}

For quantum circuits, a natural measure of complexity is the gate count. Suppose a gate set $S \subseteq \mathcal{U}(\mathcal{H})$ is given. Then, the (approximate) gate count for $U \in \mathcal{U}(\mathcal{H})$ relative to $S$, given a fixed error threshold $\delta \geq 0$, is defined by \cite{HFKEH,Haferkamp23,Nietner23}:
\begin{equation}\label{def:depth}
\begin{aligned}
        & \operatorname{Count}_{\delta}^S(U) \\
        & := \min\left\{ l \geq 1 : \| U - V_1 \cdots V_l \|_{\operatorname{op}} \leq \delta,\ V_i \in S \right\}.
\end{aligned}
\end{equation}

Any unitary $U$ can be viewed as a quantum channel $\adm_U$:
\begin{equation}\label{eqn:unitary operator}
   \adm_U(\rho) = U \rho U^{\dagger}.
\end{equation}

Suppose we can implement a circuit using a sequence of random gates, and the average gate (a mixed-unitary channel) is close to the target circuit. This motivates the convexified gate count, defined by \cite{Campbell19}:
\begin{equation}\label{def:convexified depth}
\begin{aligned}
    l^S_{\delta}(U) := \inf \big\{ l \geq 1 : \exists\ \mu \in \mathcal{P}(S^l)\ \text{such that} \\
    \left\| \adm_U - \int_{S^l} \adm_{u_1} \circ \cdots \circ \adm_{u_l}\ d\mu(u_1, \dots, u_l) \right\|_{\diamond} \leq \delta \big\},
\end{aligned}  
\end{equation}
where $\mathcal{P}(S^l)$ is the set of probability measures on $S^l = S \times \cdots \times S$, the Cartesian product of $S$ with itself $l$ times.

Note that using the well-known relation for two unitary operators $U, V$,
\begin{equation*}
    \| \adm_U - \adm_V \|_{\diamond} = 2\| U - V \|_{\operatorname{op}},
\end{equation*}
we have $l^S_{\delta}(U) \leq \operatorname{Count}^S_{\delta/2}(U)$, which follows directly by choosing the random gates as the deterministic ones in \eqref{def:convexified depth}.

\added{As a first application, we show that our notion of complexity provides a lower bound for the (convexified) gate count. A more intricate application of this idea can be found in \cite{ding2024lower}. }

\begin{theorem}\label{main:Gate count}
Given a unitary operator $U \in \mathcal{U}(\mathcal{H})$ and a resource set $S$ of unitary operators, we have
\begin{equation}\label{lower bound: exact}
C_S^{\operatorname{cb}}(\adm_U) \leq l^S_{0}(U).
\end{equation}
Moreover, if $U \in S''$ (the double commutant of $S$, defined as the operators commuting with $S'$), then for any $\delta \geq 0$,
\begin{equation}\label{lower bound: approximate}
    C_S^{\operatorname{cb}}(\adm_U) \leq l^S_{\delta}(U) + \delta\, \kappa(S),
\end{equation}
where $\kappa(S)$ is the quantum expected length defined in \eqref{def:expected length}.
\end{theorem}

Note that in most cases, $S'$ consists of scalars; thus, $S''$ is the entire matrix algebra, and the technical assumption $U \in S''$ is always satisfied. The proof of the theorem is provided in Appendix \ref{app:proof of Gate count}.

\subsection{Wasserstein Complexity}

Quantum Wasserstein complexity of order 1 for quantum channels, $C_{W_1}(\cdot)$, was first introduced in \cite{LBKJL}. We show that by choosing our resource set $S$ appropriately (e.g., Pauli gates), $C_S(\cdot)$ is equivalent to $C_{W_1}(\cdot)$. We first review the definition of Wasserstein complexity of order 1.

The underlying Hilbert space is $\mathcal{H} = (\mathbb{C}^d)^{\otimes n}$, and the Lipschitz norm is given by \cite[Eq.~(6.46)]{depalma2023quantumoptimaltransportquantum}:
\begin{equation}\label{def:Lipschitz norm Wasserstein 1}
    \| x \|_L = 2 \max_{1 \leq i \leq n} \min_{H^{(i)}} \| x - \mathbb{I}_i \otimes H^{(i)} \|_{\operatorname{op}},
\end{equation}
where the minimum is over Hermitian operators $H^{(i)}$ acting on the $n-1$ qudits excluding the $i$-th register.

The above Lipschitz norm naturally defines an Earth Mover's distance on the state space, known as the Wasserstein-1 distance (see explicit expressions in \cite{PMTL,depalma2023quantumoptimaltransportquantum}):
\begin{equation}
    W_1(\rho, \sigma) = \sup \left\{ \left| \operatorname{tr}( x (\rho - \sigma) ) \right| : \| x \|_L \leq 1,\ x = x^{\dagger} \right\}.
\end{equation}

The Wasserstein complexity of quantum channels, introduced in \cite{LBKJL}, can be rewritten as
\begin{align*}
    C_{W_1}(\Phi) & = \sup_{\rho \in \mathcal{D}(\mathcal{H})} W_1\left( \Phi(\rho), \rho \right) \\
    & = \sup_{ \| x \|_L \leq 1 } \| \Phi^*(x) - x \|_{\operatorname{op}}.
\end{align*}

\begin{theorem}\label{main:Wasserstein}
Suppose $S = \{ \sigma_i^j : 1 \leq j \leq d^2 - 1,\ 1 \leq i \leq n \}$ is the generalized Pauli gate set, where for each $1 \leq j \leq d^2 - 1$, $\sigma^j$ is a generalized Pauli operator, and for each $1 \leq i \leq n$,
\begin{equation}
    \sigma_i^j := \mathbb{I}_d^{\otimes (i-1)} \otimes \sigma^j \otimes \mathbb{I}_d^{\otimes (n - i)}.
\end{equation}
Then for any quantum channel $\Phi$,
\[
\frac{1}{2} C^{cb}_S(\Phi) \leq C^{cb}_{W_1}(\Phi) \leq \left( 2 - \frac{2}{d^2} \right) C^{cb}_S(\Phi).
\]
\end{theorem}

\noindent The proof is given in Appendix \ref{app: proof of Wasserstein}.

\subsection{Geometric Complexity}\label{sec:geometric}

In this section, we discuss another aspect of the flexibility of our resource-dependent complexity measure. In the definition \eqref{eqn:Lipschitz_norm}, we take the maximum over all possible resources, which can be viewed as an $\ell^{\infty}$-type norm. We can also introduce an $\ell^2$-type norm. Suppose $S = \{s_1, \ldots, s_m\}$; we define
\begin{equation}\label{def:Lipschitz l^2}
    ||| x |||_{S,2} := \left\| \left( \sum_{j=1}^m | [s_j, x] |^2 \right)^{1/2} \right\|_{\operatorname{op}}.
\end{equation}
We can naturally define a complexity measure based on the above Lipschitz semi-norm:
\begin{align}\label{def:p-complexity}
    C_{S,2}(\Phi) = \sup_{||| x |||_{S,2} \leq 1} \| \Phi^*(x) - x \|_{\operatorname{op}}. 
\end{align}
Our complexity measure $C_{S,2}$ provides a lower bound for the geometric complexity introduced in \cite{N1,N2,N3}.

First, we briefly review the idea of defining geometric complexity using a metric on the space of Hamiltonians. For any $U \in SU(2^n)$ (unitary operators with determinant 1), there exists a time-dependent Hamiltonian connecting the identity operator to $U$ via the following evolution:
\begin{align}\label{eqn:evolution}
  \frac{dU(t)}{dt}= -i H(t) U(t), \quad U(0) = \mathbb{I}, \quad U(1) = U.
\end{align}
We denote
\begin{align}\label{eqn:path ordering}
    U(t) = \mathcal{P} \exp\left( -i \int_0^t H(s)\, ds \right)
\end{align}
as the path-ordered unitary operator, which is the unique solution to \eqref{eqn:evolution}. The \textit{geometric complexity} is defined by
\begin{equation}
\begin{aligned}
       & C_{\operatorname{geom}}(U) = \inf\bigg\{ \int_{0}^{1} \| H(t) \|_{\operatorname{cost}} dt: \\
       & \hspace{2cm} U= U(1) = \mathcal{P} \exp\left( -i \int_0^1 H(s)\, ds \right) \bigg\}.
\end{aligned}
\end{equation}
The norm $\| H(t) \|_{\operatorname{cost}}$ encapsulates the cost metric, which must be specified based on the context or physical considerations. One particular choice given in \cite{N1,N2,N3} is as follows. Suppose $\mathfrak{H} = \{ \sigma_{\vec{k}} : \vec{k} = (k_1, \ldots, k_n),\ k_i = 0,1,2,3 \}$ is the set of tensor products of Pauli operators,
\begin{align*}
    \sigma_{\vec{k}} = \bigotimes_{i=1}^n \sigma_{k_i},
\end{align*}
where the Pauli operators are given by $\sigma_0 = \mathbb{I}_2$, $\sigma_1 = \sigma_x$, $\sigma_2 = \sigma_y$, $\sigma_3 = \sigma_z$. For any Hermitian operator $H(t)$, we have
\begin{equation}
    H(t) = \sum_{\vec{k}} \alpha_{\vec{k}}(t)\, \sigma_{\vec{k}}.
\end{equation}
We associate to each direction $\vec{k}$ a penalty factor $p_{\vec{k}} \in (0, \infty]$; then the cost metric is defined as
\begin{equation}\label{def:cost metric}
    \| H(t) \|_{\operatorname{cost}} = \sqrt{ \sum_{\vec{k}} p_{\vec{k}}\, |\alpha_{\vec{k}}(t)|^2 }.
\end{equation}

\begin{theorem}\label{main: Geometric complexity}
Suppose our resource set $S$ is given by
\begin{equation}
    S = \left\{ \frac{1}{\sqrt{p_{\vec{k}}}}\, \sigma_{\vec{k}} \ :\ \vec{k} = (k_1, \ldots, k_n),\ k_i = 0,1,2,3 \right\}.
\end{equation}
Then for any $U \in SU(2^n)$, we have
\begin{equation}\label{eq: nielsen}
    C_{S,2}(\adm_{U}) \leq C_{\operatorname{geom}}(U),
\end{equation}
with the cost metric given by \eqref{def:cost metric}.
\end{theorem}

The proof is given in Appendix \ref{app:proof of Geometric}. The above lower bound can be viewed as a resource-dependent version of the lower bound on the cost of circuits, which has been studied extensively in \cite{bu2024complexity, LBKJL}. By generalizing this framework to arbitrary Lie groups, our approach exhibits greater flexibility in the choice of the cost function and the underlying basis for the Lie algebra. We can even achieve equality via approximation using the components of the left regular representation; see \cite{araiza2025} for full details.

\section{Linear growth lower bound for closed and open systems}

In \cite{BS1, BS2}, it is conjectured that the complexity of a typical quantum circuit grows linearly with depth and saturates at a value that is exponential in the system size. In this section, we show that for our resource-dependent complexity measure, this phenomenon also occurs in both closed and open quantum systems. The saturation value, denoted as $\kappa(S)$ in our paper, depends on the system size and the resource set $S$, which can be constant, polynomial, or exponential in the system size.

\subsection{Random Circuits}\label{subsection:random circuits}

In this subsection, we construct random circuits that exhibit linear growth of Lipschitz complexity, providing a lower bound for gate count. To be precise, let us fix a resource set $S$ consisting of unitary operators, which can be one-qubit or two-qubit gates, or parallel circuits with certain geometric structures. Our random circuit model consists of concatenating random samples from $S$:

\begin{center}
\begin{quantikz}[scale=1] 
\lstick{$|\psi\rangle$} & \gate[3]{U_{1}} & \gate[3]{U_{2}} & \qw \cdots & \qw & \gate[3]{U_{l}}& \meter{}\\
\lstick{$|\psi\rangle$} &          &  & \qw \cdots & \qw & &  \meter{} \\
\lstick{$|\psi\rangle$} &          &  & \qw \cdots & \qw & &  \meter{}
\end{quantikz}
\end{center}

To achieve a rigorous analysis, we further assume that there is a spectral gap for the probability distribution $\nu$. That is, for the mixed-unitary channel induced by $\nu$, defined by
\begin{equation}
    \Phi_{\nu} := \mathbb{E}_{U \sim \nu} \adm_U = \int_S \adm_U\, \nu(dU),
\end{equation}
and for the trace-preserving conditional expectation $E_{\text{fix}}$ onto $S'$, we have
\begin{equation}\label{def:spectral gap}
    1 - \lambda_{\text{spec}} := \| \Phi_{\nu} - E_{\text{fix}} \|_2 < 1,
\end{equation}
where the $2$-norm of the super-operator is induced by the Schatten 2-norm (also known as the Frobenius norm or Hilbert-Schmidt norm):
\[
 \| \Phi_{\nu} - E_{\text{fix}} \|_2 = \sup_{\| X \|_2 \leq 1} \| \Phi_{\nu}(X) - E_{\text{fix}}(X) \|_2,
\]
and the Schatten $p$-norm of operators is given by $\| X \|_p = \left(\operatorname{tr}(|X|^p) \right)^{1/p}$, for $p > 0$.

Using spectral decomposition under the Hilbert-Schmidt norm, if $\Phi_{\nu}$ is symmetric (i.e., $\Phi_{\nu} = \Phi_{\nu}^*$), then \eqref{def:spectral gap} always holds in the finite-dimensional case. However, our analysis allows for more general mixed-unitary channels. Under the assumption \eqref{def:spectral gap}, we can show the following:

\begin{prop}\label{main:expected complexity lower bound}
    Suppose $\{ U_l \}_{l \geq 1}$ is a sequence of independent random unitaries with the same distribution $\nu$. Then for any $l \geq 1$,
    \begin{equation}
        \mathbb{E} \left[ C_S^{\operatorname{cb}}( \adm_{U_l \cdots U_1} ) \right] \geq \left( 1 - \mathcal{I}^{\operatorname{cb}}(E_{\text{fix}}) \cdot (1 - \lambda_{\text{spec}})^l \right) \kappa(S),
    \end{equation}
    where the index $\mathcal{I}^{\operatorname{cb}}(E_{\text{fix}})$ is defined in Definition \ref{def:index}.
\end{prop}

The constant $\mathcal{I}(E_{\text{fix}})$, defined for arbitrary conditional expectations, was first introduced in \cite{pimsner1986entropy}. We refer the reader to Appendix \ref{app:conditional expectation} for more details. Using the above proposition, we can obtain the following \textit{Brown-Susskind behavior} for the resource-dependent complexity of random circuits (see Table \ref{tab:table} for a summary):

\begin{theorem} \label{main: Brown-Susskind}
Suppose $\{ U_l \}_{l \geq 1}$ is a sequence of independent random unitaries with the same distribution $\nu$. For the Brown-Susskind threshold
\begin{equation}
    L = \frac{ -\log \left( 4 \mathcal{I}^{\operatorname{cb}}(E_{\text{fix}}) \right) }{ \log(1 - \lambda_{\text{spec}}) },
\end{equation}
we have for $l \leq L$,
\begin{equation}
    \mathbb{P}\left( C_S^{\operatorname{cb}}( \adm_{U_l \cdots U_1} ) \geq l \frac{ \kappa(S) }{8L} \right) \geq \frac{l}{4(l + L)}.
\end{equation}
\end{theorem}

The proofs of Proposition \ref{main:expected complexity lower bound} and Theorem \ref{main: Brown-Susskind} are given in Appendix \ref{app:linear growth}.

\added{Note that our theorem can verify the existence of quantum circuits with high complexity (exponential in the number of qubits) if $\kappa(S)$ is exponential. Compared to \cite{HFKEH}, where the author used a specific architecture, our theorem is more general in the sense that $S$ can be arbitrary. This leads to some sacrifice on the probability estimate in the sense that the probability of linear growth is not close to one.}
\begin{table*}[htbp]
\centering
\begin{tabular}{|c|c|} \hline 
\textbf{Gate Sets} & \textbf{Complexity, Threshold, Probability} \\ \hline
Sufficiently connected circuits \cite{HFKEH} &  
Exact gate count, \\ 
& Exponential, \\ 
& Linear growth with probability 1 \\ \hline
Universal gate sets \cite{Haferkamp23, chen2024incompressibility} & 
Approximate gate count, \\ 
& Exponential, \\ 
& Linear growth with probability close to 1 \\ \hline
\textbf{This Letter: Arbitrary Gate Set} &    
Resource-dependent complexity, \\ 
& Brown-Susskind threshold, \\ 
& Linear growth with positive probability \\ \hline
\end{tabular}
\caption{A summary of linear growth behavior for random circuits}
\label{tab:table}
\end{table*}

\subsection{Hamiltonian Simulation}

In this subsection, we illustrate our notion of complexity by providing a lower bound on the gate count for Hamiltonian simulation via the \textit{qDRIFT} protocol introduced in \cite{Campbell19}. The starting point is a Hamiltonian given by $H = \sum_{j=1}^m h_j H_j$, where $H_j$ are elementary components such that $\| H \|_{\operatorname{op}} \leq 1$, and $h_j$ are positive weights. We denote
\begin{equation}
    \lambda := \sum_{j=1}^m h_j.
\end{equation}

In this setting, we consider the \textit{resource set}
\begin{equation}
    S = \{ H_j \mid 1 \leq j \leq m \}.
\end{equation}
Our goal is to approximate the unitary $U(t) = \exp(i t H)$ using a product of gates from $\{ U_j(r) = \exp(i r H_j) \mid 1 \leq j \leq m,\ |r| \leq \tau \}$ up to some desired precision, where $\tau > 0$ is a predetermined constant. Note that a well-established approach is the Trotter-Suzuki formula; see \cite{Childs_2021} for a comprehensive overview of the method. Moreover, it has been shown in \cite{Campbell19, Childs_2019} that by allowing $U_j(r)$ to be randomly chosen, the number of elementary gates needed to approximate $\exp(i t H)$ can be lower than the number given by the Trotter-Suzuki formula, although the approximation is only achieved in an average sense. This motivates us to define the \textit{convexified simulation cost} using the gate set
\begin{equation}
    \mathcal{U}_\tau := \left\{ \exp(i r H_j) \mid 1 \leq j \leq m,\ |r| \leq \tau \right\}.
\end{equation}

\begin{definition}
The convexified simulation cost of $\exp(i t H)$ using gates from $\mathcal{U}_\tau$ with error $\delta > 0$ is defined by
\begin{equation}\label{def:convexified simulation cost}
\begin{aligned}
    & \overline{l}^{\mathcal{U}_\tau}_{\delta}\left( \exp(i t H) \right) := \inf \bigg\{ l \geq 1 \ \bigg|\  \exists\ \mu_l \in \mathcal{P}\left( \mathcal{U}_\tau^l \right) \ \text{such that} \\
    & \left\| \adm_{\exp(i t H)} - \int_{\mathcal{U}_\tau^l} \adm_{u_1} \cdots \adm_{u_l}\ d\mu_l(u_1, \ldots, u_l) \right\|_{\diamond} \leq \delta \bigg\},
\end{aligned}
\end{equation}
where $\mathcal{P}\left( \mathcal{U}_\tau^l \right)$ is the set of probability measures on $\mathcal{U}_\tau^l = \mathcal{U}_\tau \times \cdots \times \mathcal{U}_\tau$, the Cartesian product of $\mathcal{U}_\tau$ with itself $l$ times.
\end{definition}

Note that in \cite{Campbell19}, the author chose $\tau_0 = \frac{t \sum_j h_j}{N} \leq \tau$, where $N$ is the length of the approximating random circuit, and $\nu$ is a discrete probability measure supported on $\{ \exp(i \tau_0 H_j) \mid 1 \leq j \leq m \} \subseteq \mathcal{U}_{\tau}$, defined by $\nu\left( \{ \exp(i \tau_0 H_j) \} \right) = \frac{h_j}{\sum_k h_k}$. Then one has \cite[Eq.~(B13)]{Campbell19}:
\begin{equation}\label{ineqn:Campbell estimate}
   \left\| \Phi_{\nu}^N - \adm_{U(t)} \right\|_{\diamond} \leq 2 \frac{\lambda^2 t^2}{N} \exp\left( \frac{2 \lambda t}{N} \right),
\end{equation}
which provides an upper bound for $\overline{l}^{\mathcal{U}_\tau}_{\delta}\left( \exp(i t H) \right)$ using the probability measure $\nu_N = \nu \times \cdots \times \nu$ by setting \eqref{ineqn:Campbell estimate} less than $\delta$:
\begin{align*}
    \overline{l}^{\mathcal{U}_\tau}_{\delta}\left( \exp(i t H) \right) \leq C \frac{\lambda^2 t^2}{\delta}.
\end{align*}

Using our complexity measure, we can provide a lower bound for $\overline{l}^{\mathcal{U}_\tau}_{\delta}\left( \exp(i t H) \right)$:

\begin{theorem}\label{main:lower bound Hamiltonian simulation}
Suppose $H \neq 0$. Then for $t \leq \frac{1}{3 \| H \|_{\operatorname{op}}}$, we have
\begin{equation}
    \overline{l}^{\mathcal{U}_\tau}_{\delta}\left( \exp(i t H) \right) \geq \frac{ \lambda_H t - \delta\, \kappa(S) }{ \tau },
\end{equation}
where $\lambda_H > 0$ is given by
\begin{equation}\label{eqn:lower bound constant general}
    \lambda_H := \sup_{ x \notin S' } \frac{ \| [ H, x ] \|_{\operatorname{op}} }{ 2 ||| x |||_S }.
\end{equation}
\end{theorem}

We present the proof in Appendix \ref{app:proof of Hamiltonian simulation}. As an application, we show an explicit lower bound for the simulation cost of the Ising Hamiltonian. Using the bounds in \cite{Harrow_2017}, it is possible to generalize the argument to other local Hamiltonians.

\begin{example}
Suppose we have an Ising Hamiltonian in the $n$-qubit system given by
\begin{equation}\label{def:Ising Hamiltonian}
    H = \sum_{1 \leq k \neq l \leq n} \gamma_{kl}\, \sigma_z^k \sigma_z^l + \sum_{j=1}^n \alpha_j\, \sigma_z^j + \sum_{s=1}^n \beta_s\, \sigma_x^s,
\end{equation}
where $\gamma_{kl}, \alpha_j, \beta_s \in \mathbb{R}$, and $\sigma^j$ denotes the Pauli operator $\sigma$ acting on the $j$-th qubit. The resource set is given by
\begin{align*}
    S = \left\{ \sigma_z^k \sigma_z^l,\ \sigma_z^j,\ \sigma_x^s \mid \forall\ k, l, j, s \right\}.
\end{align*}
Then we have
\begin{align*}
    \lambda_H \geq \frac{ \left| \sum_{k \neq l} \gamma_{kl} + \frac{1}{\sqrt{2}} \sum_j \alpha_j - \frac{1}{\sqrt{2}} \sum_s \beta_s \right| }{ 4 }.
\end{align*}
In particular, if the coefficients are positive and $\gamma_{kl}, \alpha_j, \beta_s = \Theta(1)$, then $\lambda_H \geq \Omega(n^2)$.
\end{example}

\begin{proof}
We use the expression \eqref{eqn:lower bound constant general}. To obtain a lower bound, we choose a special test operator:
\begin{equation}
    X = \sum_{j=1}^n \sigma_y^j.
\end{equation}
We have
\begin{align*}
    [H, X] &= \sum_{1 \leq k \neq l \leq n} \gamma_{kl} \left( [ \sigma_z^k \sigma_z^l, \sigma_y^k ] + [ \sigma_z^k \sigma_z^l, \sigma_y^l ] \right) \\
    & \quad + \sum_{j=1}^n \alpha_j [ \sigma_z^j, \sigma_y^j ] + \sum_{s=1}^n \beta_s [ \sigma_x^s, \sigma_y^s ] \\
    &= -2i \bigg( \sum_{1 \leq k \neq l \leq n} \gamma_{kl} ( \sigma_x^k \sigma_z^l + \sigma_z^k \sigma_x^l ) \\
    & \hspace{1cm}+ \sum_j \alpha_j \sigma_x^j - \sum_s \beta_s \sigma_z^s \bigg).
\end{align*}
Now define $\ket{\phi} = \cos \left( \frac{\pi}{8} \right) \ket{0} + \sin \left( \frac{\pi}{8} \right) \ket{1}$. We have
\begin{align*}
    & \bra{\phi} \sigma_x \ket{\phi} = \sin \left( \frac{\pi}{4} \right) = \frac{1}{\sqrt{2}}, \\
    & \bra{\phi} \sigma_z \ket{\phi} = \cos \left( \frac{\pi}{4} \right) = \frac{1}{\sqrt{2}}.
\end{align*}
Therefore,
\begin{align*}
    \| [ H, X ] \|_{\operatorname{op}} &\geq \left| \bra{\phi}^{\otimes n} [ H, X ] \ket{\phi}^{\otimes n} \right| \\
    &= 2 \left| \sum_{k \neq l} \gamma_{kl} + \frac{1}{\sqrt{2}} \sum_j \alpha_j - \frac{1}{\sqrt{2}} \sum_s \beta_s \right|.
\end{align*}
Moreover, it is routine to check that
\begin{align*}
    \sup_{ s \in S } \| [ s, X ] \|_{\operatorname{op}} &= \max \left\{ \| [ \sigma_z^k \sigma_z^l, \sigma_y^k ] + [ \sigma_z^k \sigma_z^l, \sigma_y^l ] \|_{\operatorname{op}} \right\} \\
    &= 2 \left\| \sigma_x \otimes \sigma_z + \sigma_z \otimes \sigma_x \right\|_{\operatorname{op}} = 4.
\end{align*}
Therefore, the conclusion is derived as follows:
\begin{align*}
    \lambda_H &= \sup_{ x \notin S' } \frac{ \| [ H, x ] \|_{\operatorname{op}} }{ 2 ||| x |||_S } \geq \frac{ \| [ H, X ] \|_{\operatorname{op}} }{ 2 ||| X |||_S } \\
    &\geq \frac{ \left| \sum_{k \neq l} \gamma_{kl} + \frac{1}{\sqrt{2}} \sum_j \alpha_j - \frac{1}{\sqrt{2}} \sum_s \beta_s \right| }{ 4 }.
\end{align*}
\end{proof}

In practice, to simulate $\exp(i t H)$, we first simulate the small-time evolution $\exp(i \Delta t H)$ and iterate it $\lceil t / \Delta t \rceil$ times. Our accessible gate set is defined up to a small-time threshold $\tau = \Delta t$. Our theorem demonstrates that for the Ising Hamiltonian defined in \eqref{def:Ising Hamiltonian}, where all coefficients are equal to the same constant, each small-time simulation requires at least $\Omega(n^2)$ gates, even when allowing random gates and approximation in the average sense.

\subsection{Open Systems}

Our notion can be naturally generalized to open systems since the complexity measures are defined for quantum channels. Suppose we have a purely dissipative quantum Markov semigroup \( T_t = \exp(tL) \), where
\[
Lx = \sum_{j=1}^m V_j x V_j^{\dagger} - \frac{1}{2} \left( V_j^{\dagger} V_j x + x V_j^{\dagger} V_j \right)
\]
is a Lindbladian generator (see \cite{lindblad1976generators}). We now use the \textit{jump operators} as the \textit{resource set} \( S = \bigcup_{j=1}^m \{ V_j, V_j^{\dagger} \} \) and we assume $T_t^*\circ E_{\text{fix}} = E_{\text{fix}}$. Using elementary properties of Lipschitz complexity, an immediate consequence is
\[
C_S^{\operatorname{cb}}(T_t) \leq \kappa(S) \| L \|_{\diamond} t.
\]

To obtain a lower bound, we use a \textit{mixing time} argument. Recall that the mixing time of the semigroup \( T_t \) is defined by
\begin{equation}
t_{\operatorname{mix}}(\varepsilon) := \inf \left\{ t > 0 : \| T_t - E_{\text{fix}}^* \|_{\diamond} \leq \varepsilon \right\}.
\end{equation}
If the Lindbladian generator is GNS symmetric (or satisfies the $\sigma$-detailed balance condition), the mixing time is finite. Using a similar argument as in Proposition \ref{main:expected complexity lower bound}, we can show the following:

\begin{theorem}\label{main:open system}
The lower bound estimate of \( C_S^{\operatorname{cb}}(T_t) \) grows linearly before \( t_{\operatorname{mix}}(\varepsilon) \) for any \( \varepsilon > 0 \) and is comparable to \( \kappa(S) \) after \( t_{\operatorname{mix}}(\varepsilon) \).
\end{theorem}

Using standard tools in optimal transport theory, we can provide generic upper estimates for \( \kappa(S) \) when \( \{ V_j \}_{j=1}^m \) form the jump operators of a Lindbladian generator with $\sigma$-detailed balance. In particular, using Corollary 6.8 in \cite{Fisher}, we obtain
\begin{align}\label{bound:CLSI}
    \kappa(S) \leq c_0\, \operatorname{CLSI}(L)^{-1/2} \sqrt{m}\, \sqrt{ \log \mathcal{I}^{\operatorname{cb}}(E_{\text{fix}}) },
\end{align}
where \( \operatorname{CLSI}(L) \) denotes the constant in the Logarithmic Sobolev Inequality for the Lindbladian \( L \) and the number $m$ of jump operators appears when we compare $|||\cdot|||_S$ and $|||\cdot|||_{S,2}$ introduced in \eqref{def:Lipschitz l^2}.

By using Pauli gates as the jump operators of the Lindbladian generator (see Section \ref{Calculation}), we recover the upper and lower qubit bounds obtained in \cite{LBKJL}. The above technique can be generalized to provide a lower-bound framework for the simulation cost of open quantum systems in general cases; see \cite{ding2024lower} for more details.

\medskip

\section{Conclusion}

Our flexible notion of complexity for quantum channels falls into the axiomatic framework from \cite{LBKJL}. By suitably choosing the resource set \( S \), we can establish a lower bound for the complexity of random circuits and continuous-time evolutions which is only known in restricted settings before. Previous lower bound trick is to reduce the problem to computing parity function \cite{Berry_2015, Zhang_2024}, and our lower bound is based on quantum optimal transport theory. The flexibility in the choice of \( S \) allows us to explore different regimes of complexity, which can be constant, polynomial in the number of qubits, or exponential in the number of qubits. We leave the discussion about the relation to non-stabilizerness for future work, as discussed in \cite{bu2024complexity}.

Finally, the linear aspect of the Brown-Susskind conjecture \cite{BS1,BS2} is confirmed in our context, up to a threshold given by the mixing time, which is determined by the geometric properties of the resource set.

\bibliographystyle{marcotomPB}
\bibliography{cost}

\appendix
\onecolumngrid
\section{Non-commutative conditional expectations and their index}\label{app:conditional expectation}

In this paper, we assume that $S'$ is a $*$-algebra. This means that for any operators $X, Y \in S'$, we have $XY \in S'$ and for any $X \in S'$, we have $X^{\dagger} \in S'$. In the finite-dimensional setting, $\|X^\dagger X\|_{op} = \|X\|_{op}^2$, so $S'$ is automatically a $C^*$-algebra.

In the finite-dimensional case, let $\mc N_{fix} \subseteq \mc B(\mc H)$ be a $C^*$-subalgebra of $\mc B(\mc H)$. A conditional expectation onto $\mc N_{fix}$ is defined as follows:

\begin{definition}\label{def:conditional expectation}
A unital, completely positive map $E_{\text{fix}}: \mc B(\mc H) \to \mc N_{fix}$ is called a \textbf{conditional expectation} if:
\begin{itemize}
    \item for all $a \in \mc N_{fix}$, $E_{\text{fix}}(a) = a$.
    \item for all $a, b \in \mc N_{fix}$ and $X \in \mc B(\mc H)$, $E_{\text{fix}}(a X b) = a E_{\text{fix}}(X) b$.
\end{itemize}
\end{definition}

Note that the dual map $E_{\text{fix}}^*$ is trace-preserving, making it a quantum channel. Recall that a finite-dimensional $C^*$-algebra is unitarily equivalent to 
\[
\bigoplus_{i=1}^n \mb B(\mc H_i) \otimes \mb C \cdot \mb I_{\mc K_i}, \quad \mc H = \bigoplus_{i=1}^n \mc H_i \otimes \mc K_i.
\]
Without loss of generality, we can assume $\mc N_{fix} = \bigoplus_{i=1}^n \mb B(\mc H_i) \otimes \mb C \cdot \mb I_{\mc K_i}$. Denote $P_i$ as the projection of $\mc H$ onto $\mc H_i \otimes \mc K_i$. Then, there exists a family of density operators $\tau_i \in \mc D(\mc K_i)$ such that
\begin{align}\label{form of conditional expectation}
& E_{\text{fix}}(X) = \bigoplus_{i=1}^n \tr_{\mc K_i} \big(P_i X P_i (\mb I_{\mc H_i} \otimes \tau_i)\big) \otimes \mb I_{\mc K_i}, \\
& E_{\text{fix}}^*(\rho) = \bigoplus_{i=1}^n \tr_{\mc K_i} \big(P_i \rho P_i\big) \otimes \tau_i,
\end{align}
where $\tr_{\mc K_i}$ denotes the partial trace with respect to $\mc K_i$. It is easy to see that a state $\sigma$ is invariant under $E_{\text{fix}}$, i.e., $E_{\text{fix}}^*(\sigma) = \sigma$, if and only if 
\[
\sigma = \bigoplus_{i=1}^n p_i \sigma_i \otimes \tau_i,
\]
for some density operators $\sigma_i \in \mc D(\mc H_i)$ and a probability distribution $\{p_i\}_{i=1}^n$. The conditional expectation is completely determined if an invariant state is specified, and we denote it as $E_{\text{fix},\sigma}$. We drop the index $\sigma$ when it is clear from the context. Note that the invariant state is not unique, and the flexibility comes from different choices of $\sigma_i$ and the probability distribution $\{p_i\}_{i=1}^n$.

Now, let us introduce the index, which quantifies the "size" of a $C^*$-algebra.

\begin{definition}\label{def:index}
For a $C^*$-algebra $\mc N_{fix}$ and a conditional expectation $E_{\text{fix}}$ onto $\mc N_{fix}$, the (completely bounded) index $\mc I(E_{\text{fix}})$ (respectively $\mc I^{cb}(E_{\text{fix}})$) is defined as
\begin{align}
& \mc I(E_{\text{fix}}) := \inf \{c > 0 \mid \rho \leq c E_{\text{fix}}^*(\rho) \text{ for all states } \rho \}, \\
& \mc I^{cb}(E_{\text{fix}}) := \sup_{n \in \mb N} \mc I(E_{\text{fix}} \otimes \id_{\mb M_n}).
\end{align}
\end{definition}

We illustrate the calculation by choosing the trace-preserving conditional expectation $E_{\text{fix},\tr}$, i.e., we take $\tau_i = \frac{1}{d_{\mc K_i}} \mb I_{\mc K_i}$ in \eqref{form of conditional expectation}. The following properties of the index are useful (see \cite{GJL20entropy} for proofs):

\begin{prop}\label{Properties I:index}
Assume $\mc N_{fix}$ has the form
\[
\mc N_{fix} = \bigoplus_{i=1}^n \mb B(\mc H_i) \otimes \mb C \cdot \mb I_{\mc K_i}, \quad \mc H = \bigoplus_{i=1}^n \mc H_i \otimes \mc K_i.
\]
Then, the index for the trace-preserving conditional expectation is given by
\begin{align}
& \mc I(E_{\text{fix},\tr}) = \sum_{i=1}^n \min\{d_{\mc H_i}, d_{\mc K_i}\} d_{\mc K_i}, \\
& \mc I^{cb}(E_{\text{fix},\tr}) = \sum_{i=1}^n d_{\mc K_i}^2. \label{index:ergodic}
\end{align}
\end{prop}

\section{Proof of Theorem \ref{main:Gate count}} \label{app:proof of Gate count}
\begin{proof}[Proof of Theorem \ref{main:Gate count}]
For the proof of \eqref{lower bound: exact}, we assume $l = l_0^S(U) < +\infty$ (the case $l = +\infty$ is trivial). Then, there exists a probability measure $\mu^l \in \mc P(S^l)$ such that  
\[
    \adm_U = \int_{S^l} \adm_{u_1}\cdots \adm_{u_l} \, d\mu^l(u_1,\ldots, u_l). 
\]
Using triangle inequalities, we have for any $x \in \mc B(\mc H)$:
\[
    \|\adm_U^*(x) - x\|_{op} \le \int_{S^l} \|\adm_{u_1}\cdots \adm_{u_l}(x) - x\|_{op} \, d\mu^l(u_1,\ldots, u_l).
\]
This simplifies as follows:
\[
    \|\adm_U^*(x) - x\|_{op} = \int_{S^l} \|[u_1\cdots u_l, x]\|_{op} \, d\mu^l(u_1,\ldots, u_l).
\]
Using the Leibniz rule for commutators, $[AB,C] = A[B,C] + [A,C]B$, and triangle inequalities,
\[
    \|[u_1\cdots u_l, x]\|_{op} \le \|u_l [u_{l-1}\cdots u_1, x]\|_{op} + \|[u_l, x] u_{l-1}\cdots u_1\|_{op}.
\]
Thus:
\[
    \|\adm_U^*(x) - x\|_{op} \le \int_{S^l} \|[u_l, x]\|_{op} + \|[u_{l-1}\cdots u_1, x]\|_{op} \, d\mu^l(u_1,\ldots, u_l).
\]
Inductively, this yields:
\begin{equation}\label{Lebniz argument}
        \|\adm_U^*(x) - x\|_{op} \le \int_{S^l} \sum_{j=1}^l \|[u_j, x]\|_{op} \, d\mu^l(u_1,\ldots, u_l) \le l \sup_{u \in S} \|[u, x]\|_{op} = l |||x|||_S.
\end{equation}

Taking the supremum over all $x \in \mc B(\mc H)$ with $|||x|||_S \le 1$, we obtain $C_S(\adm_U) \le l_0^S(U)$. The same argument applies if we replace the elements of $\mc B(\mc H)$ with those of $\mb B(\mc H \otimes \mc H_d)$, giving $C_S^{cb}(\adm_U) \le l_0^S(U)$.

For the proof of \eqref{lower bound: approximate}, assume $l = l_{\delta}^S(U) < +\infty$. Then, there exists a probability measure $\mu^l \in \mc P(S^l)$ such that 
\[
    \left\|\adm_U - \int_{S^l} \adm_{u_1}\cdots \adm_{u_l} \, d\mu^l(u_1,\ldots, u_l)\right\|_{\diamond} \le \delta.
\]
Define $\Psi_l = \int_{S^l} \adm_{u_1}\cdots \adm_{u_l} \, d\mu^l(u_1,\ldots, u_l)$. Since $U \in S''$, $U$ commutes with any element in $S'$, and we have 
\[
    \adm_U^* \circ E_{\text{fix}} = \Psi_l^* \circ E_{\text{fix}} = E_{\text{fix}},
\]
where $E_{\text{fix}}$ is the conditional expectation onto $S'$. For any $x \in \mc B(\mc H)$, this implies:
\[
    \|\adm_U^*(x) - x\|_{op} \le \|\adm_U^*(x) - \Psi_l^*(x)\|_{op} + \|\Psi_l^*(x) - x\|_{op} \le \|(\adm_U^* - \Psi_l^*)(E_{\text{fix}}(x) - x)\|_{op} + \|\Psi_l^*(x) - x\|_{op}
\]
Using operator norms and previous argument \eqref{Lebniz argument},
\[
    \|\adm_U^*(x) - x\|_{op} \le \|\adm_U^* - \Psi_l^*\|_{\infty} \|E_{\text{fix}}(x) - x\|_{op} + l |||x|||_S.
\]
Here, the first term follows from the norm equivalence:
\[
    \|T^*\|_{\infty} = \sup_{\|x\|_{op} \le 1} \|T^*(x)\|_{op} = \|T\|_{1} = \sup_{\|y\|_{1} \le 1} \|T(y)\|_{1}.
\]
Taking the supremum over all $x \in \mc B(\mc H)$ and replacing elements in $\mc B(\mc H)$ with those in $\mb B(\mc H \otimes \mc H_d)$, we get:
\[
    C_S^{cb}(\adm_U) \le l_{\delta}^S(U) + \|\adm_U - \Psi_l\|_{\diamond} C_S^{cb}(E_{\text{fix}}^*) \le l_{\delta}^S(U) + \delta \kappa(S).
\]
\end{proof}

\section{Proof of Theorem \ref{main:Wasserstein}}\label{app: proof of Wasserstein}
We first review the basics of generalized Pauli gates in a $d$-dimensional quantum system $\mc H_d = \text{span}\{\ket{0}, \ket{1}, \cdots, \ket{d-1}\}$. The generalized Pauli-$X$ and Pauli-$Z$ operators are given by 
\[
    \sigma_{X,d} = \sum_{j=0}^{d-1} \ketbra{j+1}{j}, \quad \sigma_{Z,d} = \sum_{j=0}^{d-1} \exp\left(\frac{2\pi i}{d} j\right) \ketbra{j}.
\]
The entire set of generalized Pauli operators is given by
\[
    \{\sigma^j\}_{1 \le j \le d^2} = \{\sigma_{X,d}^t \sigma_{Z,d}^s\}_{0 \le t,s \le d-1}.
\]
A key observation is that a quantum channel with Kraus representation given by the set of generalized Pauli operators is actually a replacer channel. This phenomenon holds as long as the Kraus operators form an orthonormal basis \cite[Lemma 4.1]{junge2022stability}:
\[
    \frac{1}{d^2} \sum_{j=1}^{d^2} (\sigma^{j})^{\dagger} x \sigma^j = \tr(x) \frac{\mb I_d}{d}, \quad x \in \mc B(\mc H_d).
\]
Applying this operation locally, for any $1 \le i \le n$, we have
\begin{equation}\label{eqn:local conditional expectation}
  \mc E_i(X) := \frac{1}{d^2} \sum_{j=1}^{d^2} (\sigma^{j}_i)^{\dagger} X \sigma^j_i = \mb I_i \otimes \frac{1}{d}\tr_i(X), \quad X \in \mb B(\mc H_d^{\otimes n}),
\end{equation}
where $\mb I_i$ denotes the identity operator on $i$-th system. 

Using the above discussion, we show that the minimization $\min_{H^{(i)}} \|X - \mb{I}_i \otimes H^{(i)}\|_{op}$ is almost achieved at $\mc E_i(X)$, which implies that our Lipschitz norm is comparable to the Lipschitz constant of the Wasserstein-1 distance when we choose our resource set as the generalized Pauli gate set. 

\begin{lemma}
    Suppose $S = \{\sigma_i^j : 1 \le j \le d^2, 1 \le i \le n\}$ is the generalized Pauli gate set. Then for any $X \in \mb B(\mc H_d^{\otimes n})$, we have 
    \[
        \frac{1}{2}|||X|||_S \le \|X\|_L \le \left(2 - \frac{2}{d^2}\right)|||X|||_S.
    \]
\end{lemma}

\begin{proof}
Fix $1 \le i \le n$. Following the above discussion, we aim to show that $\min_{H^{(i)}} \|X - \mb{I}_i \otimes H^{(i)}\|_{op}$ and $\sup_{1 \le j \le d^2} \|[\sigma_i^j, X]\|_{op}$ are both comparable to $\|X - \mc E_i(X)\|_{op}$. Maximizing over $i$ will then complete the proof.

\noindent \textbf{Step I:} $\|X - \mc E_i(X)\|_{op} \sim \min_{H^{(i)}} \|X - \mb{I}_i \otimes H^{(i)}\|_{op}$. \\
It is clear that $\min_{H^{(i)}} \|X - \mb{I}_i \otimes H^{(i)}\|_{op} \le \|X - \mc E_i(X)\|_{op}$ by choosing $H^{(i)} = \frac{1}{d}\tr_i(X)$. For the reverse inequality, for any $H^{(i)} \in \mb B(\mc H_d^{\otimes (n-1)})$ that avoids the $i$-th register, we have
\[
    \|X - \mc E_i(X)\|_{op} = \|X - \mc E_i(X) + \mb{I}_i \otimes H^{(i)} - \mb{I}_i \otimes H^{(i)}\|_{op}.
\]
This expands as
\[
    \|X - \mc E_i(X)\|_{op} = \|X - \mb{I}_i \otimes H^{(i)} - \mc E_i\left(X - \mb{I}_i \otimes H^{(i)}\right)\|_{op} \le 2 \|X - \mb{I}_i \otimes H^{(i)}\|_{op},
\]
where we use the contraction property of the quantum channel $\mc E_i$, i.e., $\|\mc E_i\|_{op} \le 1$. In summary, we have shown
\begin{equation}\label{eqn: Intermediate 1}
    \frac{1}{2}\|X - \mc E_i(X)\|_{op} \le \min_{H^{(i)}} \|X - \mb{I}_i \otimes H^{(i)}\|_{op} \le \|X - \mc E_i(X)\|_{op}.
\end{equation}

\noindent \textbf{Step II:} $\|X - \mc E_i(X)\|_{op} \sim \sup_{1 \le j \le d^2} \|[\sigma_i^j, X]\|_{op}$. \\
Using \eqref{eqn:local conditional expectation}, we have 
\[
    \|X - \mc E_i(X)\|_{op} = \left\|\frac{1}{d^2} \sum_{j=1}^{d^2} \left(X - (\sigma^{j}_i)^{\dagger} X \sigma^j_i\right)\right\|_{op} \le \frac{d^2 - 1}{d^2} \sup_{1 \le j \le d^2} \|[\sigma_i^j, X]\|_{op},
\]
where we use the fact that one of the $\sigma_i^j$ is the identity operator. For the reverse inequality, for any $1 \le j \le d^2$, we have 
\[
    \|X - (\sigma^{j}_i)^{\dagger} X \sigma^j_i\|_{op} = \|X - \mc E_i(X) + \mc E_i(X) - (\sigma^{j}_i)^{\dagger} X \sigma^j_i\|_{op} \le 2 \|X - \mc E_i(X)\|_{op}.
\]
In summary, we obtain
\[
    \frac{d^2}{d^2 - 1}\|X - \mc E_i(X)\|_{op} \le \sup_{j} \|[\sigma_i^j, X]\|_{op} \le 2 \|X - \mc E_i(X)\|_{op}.
\]
Combining the inequality \eqref{eqn: Intermediate 1}, we conclude
\[
   \min_{H^{(i)}} \|X - \mb{I}_i \otimes H^{(i)}\|_{op} \in  \left[\frac{1}{4}, \frac{d^2 - 1}{d^2}\right] \cdot \sup_{j} \|[\sigma_i^j, X]\|_{op}.
\]
Maximizing over $i$ and using the definition of $\|\cdot\|_L$, we finally have 
\[
    \frac{1}{2} |||X|||_S \le \|X\|_L \le \left(2 - \frac{2}{d^2}\right)|||X|||_S.
\]
\end{proof}

Using the definition of the two complexity measures, the comparison is a direct corollary of the above lemma. Note that the above argument works exactly the same if we replace the elements in $\mc B(\mc H)$ by the elements in $\mb B(\mc H \otimes \mc H_d)$, so the $cb$ version also holds.

\section{Proof of Theorem \ref{main: Geometric complexity}} \label{app:proof of Geometric}
\begin{proof}
    Suppose the time-dependent Hamiltonian $H(t)$ achieves the infimum of $C_{geom}(U)$, and let $U(t) = \mc P \exp\left(-i \int_0^t H(s) \, ds\right)$ with $U(1) = U$. For any observable $X$, we aim to bound $\|U(1)^{\dagger}X U(1) - X\|_{op}$. The derivative at $t \in [0,1]$ is given by 
    \[
        \frac{d}{dt}U(t)^{\dagger}X U(t) = iU(t)^{\dagger} H(t)X U(t) - iU(t)^{\dagger}X H(t) U(t) = i U(t)^{\dagger} [H(t), X] U(t).
    \]
    Therefore, we have 
    \[
        \|U(1)^{\dagger}X U(1) - X\|_{op} = \left\| \int_0^1 \frac{d}{dt}U(t)^{\dagger}X U(t) \, dt \right\|_{op} \le \int_0^1 \|[H(t), X]\|_{op} \, dt.
    \]
    Note that 
    \[
        H(t) = \sum_{\vec{k}} \alpha_{\vec{k}}(t) \sigma_{\vec{k}} = \sum_{\vec{k}} \sqrt{p_{\vec k}} \alpha_{\vec{k}}(t) \frac{\sigma_{\vec{k}}}{\sqrt{p_{\vec k}}},
    \]
    plugging it in, we get 
    \[
        \|U^{\dagger}X U - X\|_{op} \le \int_0^1 \|[H(t), X]\|_{op} \, dt = \int_0^1 \left\|\sum_{\vec k}\sqrt{p_{\vec k}}\alpha_{\vec{k}}(t) \left[\frac{\sigma_{\vec{k}}}{\sqrt{p_{\vec k}}}, X\right]\right\|_{op} \, dt.
    \]
    Using the triangle inequality and Cauchy-Schwarz, we further obtain:
    \[
        \int_0^1 \left\|\sum_{\vec k}\sqrt{p_{\vec k}}\alpha_{\vec{k}}(t) \left[\frac{\sigma_{\vec{k}}}{\sqrt{p_{\vec k}}}, X\right]\right\|_{op} \, dt 
        \le \int_0^1 \sqrt{\sum_{\vec k} p_{\vec k}|\alpha_{\vec{k}}(t)|^2} \, dt \cdot \left\|\left(\sum_{\vec k} \left| \left[\frac{\sigma_{\vec{k}}}{\sqrt{p_{\vec k}}}, X \right] \right|^2\right)^{1/2}\right\|_{op}.
    \]
    Using the definition of the semi-norm $|||X|||_{S,2}$, we have:
    \[
        \|U^{\dagger}X U - X\|_{op} \le |||X|||_{S,2} \int_0^1 \|H(t)\|_{cost} \, dt.
    \]
    By the definition of $C_{geom}(U)$ in \eqref{def:p-complexity}, we conclude the proof:
    \[
        C_{S,2}(\adm_U)\le C_{geom}(U) = \int_0^1 \|H(t)\|_{cost} \, dt.
    \]
\end{proof}

\section{Linear growth for random circuits}\label{app:linear growth}
\textit{Conditional norm trick.}---To prove Proposition \ref{main:expected complexity lower bound}, we exploit the fact that for sufficiently large $l$, $\Phi_{\nu}^l$ approximates $E_{\text{fix}}$ closely in the presence of a spectral gap. To establish a lower bound, we employ the concept of \textit{return time}, systematically studied in \cite{GJLL22}. We begin by reviewing the \textit{conditional norm trick}, which links the spectral gap $\|\Phi^l_{\nu} - E_{\text{fix}}\|_2$ to the completely positive (cp) order: 
\[
(1-\varepsilon) E_{\text{fix}} \le_{cp} \Phi^l_{\nu} \le_{cp} (1+\varepsilon) E_{\text{fix}}.
\]
The cp order between two super-operators $T_1$ and $T_2$ is defined as $T_1 \le_{cp} T_2$ if $T_2 - T_1$ is completely positive.

To establish this connection, we introduce the non-commutative conditional $L_p^q$ norm. For a detailed treatment, refer to \cite{JPmem}. Here, we focus on a specific case. For a $C^*$-algebra $\mc N_{\text{fix}} \subseteq \mc B(\mc H)$, the $L_{\infty}^1$ norm is defined as:
\[
    \|X\|_{L_{\infty}^1(\mc N_{\text{fix}})} = \sup_{a,b \in \mc N_{\text{fix}}, \|a\|_2,\|b\|_2 \le 1} \|aXb\|_1.
\]
The induced norm for super-operators is given by:
\[
    \|T\|_{1\to \infty} := \sup_{\|X\|_{L_{\infty}^1(\mc N_{\text{fix}})} \le 1} \|T(X)\|_{op}.
\]
The completely bounded (cb) version is defined as:
\[
    \|T\|_{1\to \infty,cb} := \sup_{n \ge 1}\sup_{\|X_n\|_{L_{\infty}^1(\mb M_n \otimes \mc N_{\text{fix}})} \le 1} \|(\id_{\mb M_n}\otimes T)(X_n)\|_{op}.
\]

One application of the $L_{\infty}^1$ norm is its ability to recover the index of trace-preserving conditional expectations as defined in Definition \ref{def:index} (see \cite[Theorem 3.9]{GJL20entropy} for details):
\[
    \mc I(E_{\text{fix}}) = \|id_{\mc B(\mc H)}\|_{1\to \infty}, \quad \mc I^{cb}(E_{\text{fix}}) = \|id_{\mc B(\mc H)}\|_{1\to \infty,cb}.
\]
Another significant property is the connection between the $L_{\infty}^1$ norm, the spectral gap, and the cp order. Recall that a map $T: \mc B(\mc H) \to \mc B(\mc H)$ is called an $\mc N_{\text{fix}}$-bimodule map if:
\[
    T(axb) = aT(x)b, \quad \forall a,b \in \mc N_{\text{fix}}, x \in \mc B(\mc H).
\]
The following properties, derived in \cite[Lemma 3.14, 3.15]{Fisher}, establish the link between the spectral gap and the cp order:
\begin{prop}\label{main:connect spetral gap and cp order}
Let $\Phi:\mc B(\mc H)\to \mc B(\mc H)$ be an $\mc N_{\text{fix}}$-bimodule map. Then:
\[
    \|\Phi\|_{1\to \infty,cb} \le \mc I^{cb}(E_{\text{fix}}) \|\Phi\|_2.
\]
Moreover, if $\Phi$ is unital and completely positive, then for $\varepsilon \in (0,1)$, $\|\Phi - E_{\text{fix}}\|_{1\to \infty, cb} \le \varepsilon$ if and only if:
\[
    (1-\varepsilon)E_{\text{fix}}\le_{cp} \Phi \le_{cp} (1+\varepsilon)E_{\text{fix}}.
\]
\end{prop}

Now we are ready to prove the main results presented in Section \ref{subsection:random circuits}:
\begin{proof}[Proof of Proposition \ref{main:expected complexity lower bound}]
    Recall that $ \|\Phi_{\nu} - E_{\text{fix}}\|_2 = 1 - \lambda_{spec}$. For any $l\ge 1$, denote $\Phi^l = \Phi \circ \cdots \circ \Phi$ as the $l$-composition of $\Phi$, we have
    \begin{align*}
        \|\Phi^l_{\nu} - E_{\text{fix}}\|_2 = \|(\Phi_{\nu} - E_{\text{fix}})^l\|_2 \le (1 - \lambda_{spec})^l.
    \end{align*}
    where we use the fact that for trace-preserving conditional expectation $E_{\text{fix}}$, we have $E_{\text{fix}} \circ \Phi_{\nu} = \Phi_{\nu} \circ E_{\text{fix}} = E_{\text{fix}}$. Then use Proposition \ref{main:connect spetral gap and cp order}, we have
    \begin{align*}
        \|\Phi_{\nu}^l - E_{\text{fix}}\|_{1\to \infty, cb} \le \mc I^{cb}(E_{\text{fix}}) (1 - \lambda_{spec})^l.
    \end{align*}
    which implies
    \begin{align*}
        \Phi_{\nu}^l \le_{cp} (1 + \mc I^{cb}(E_{\text{fix}}) (1 - \lambda_{spec})^l) E_{\text{fix}}.
    \end{align*}
    Denote $\varepsilon = \mc I^{cb}(E_{\text{fix}}) (1 - \lambda_{spec})^l$, and note that $\Phi_{\nu}$ and $E_{\text{fix}}$ are both unital quantum channels, thus there exists another quantum channel $\mc R$ such that 
    \begin{align*}
        E_{\text{fix}} = \frac{1}{1+\varepsilon} \Phi_{\nu}^l + \frac{\varepsilon}{1+\varepsilon} \mc R.
    \end{align*}
    Then using convexity and the upper bound of $C_S^{cb}$, see Lemma \ref{lemma:elementary property}, we have 
    \begin{align*}
        \kappa(S):= C_S^{cb}(E_{\text{fix}}) & \le \frac{1}{1+\varepsilon} C_S^{cb}(\Phi_{\nu}^l) + \frac{\varepsilon}{1+\varepsilon} C_S^{cb}(\mc R) \le \frac{1}{1+\varepsilon} C_S^{cb}(\Phi_{\nu}^l) + \frac{2\varepsilon}{1+\varepsilon} \kappa(S),
    \end{align*}
    which implies $ C_S^{cb}(\Phi_{\nu}^l) \ge (1-\varepsilon)\kappa(S)$. Finally, using the fact that $\mb E(\adm_{U_l \cdots U_1}) = \Phi_{\nu}^l$ and convexity, we have
    \begin{align*}
        \mb E C_S^{cb}(\adm_{U_l \cdots U_1}) & \ge C_S^{cb}(\mb E(\adm_{U_l \cdots U_1})) = C_S^{cb}(\Phi_{\nu}^l) \ge (1-\varepsilon)\kappa(S) = \big(1-\mc I^{cb}(E_{\text{fix}}) (1 - \lambda_{spec})^l\big)\kappa(S).
    \end{align*}
\end{proof}
From Proposition \ref{main:expected complexity lower bound}, it is easy to see that for any $\varepsilon \in (0,1)$, define 
\begin{equation}
 L_0(\varepsilon) := \frac{\log \varepsilon - \log \mc I^{cb}(E_{\text{fix}})}{\log(1 - \lambda_{spec})},
\end{equation}
for any $l\ge L_0(\varepsilon)$, $\mb E C^{cb}_S(\adm_{U_l \cdots U_1}) \ge (1-\varepsilon) \kappa(S)$. Using a standard probabilistic argument, we can prove Theorem \ref{main: Brown-Susskind}:

\begin{proof}[Proof of Theorem \ref{main: Brown-Susskind}]
   First, we claim that for any $l \ge L_0(\varepsilon)$ and any sufficiently small $\delta > 0$, the following holds:
\begin{equation}\label{ineqn: large time positive probability}
\mb P(C^{cb}_S(\adm_{U_l \cdots U_1}) \ge \delta \kappa(S)) \ge \frac{1-\varepsilon-\delta}{2}.
\end{equation}
To see this, observe that for any $l \ge L_0(\varepsilon)$,
\begin{align*}
&(1-\varepsilon)\kappa(S) \le \mb E C^{cb}_S(\adm_{U_l \cdots U_1}) \\
&= \mb E \big[C^{cb}_S(\adm_{U_l \cdots U_1}); C^{cb}_S(\adm_{U_l \cdots U_1}) \ge \delta \kappa(S) \big]  + \mb E \big[C^{cb}_S(\adm_{U_l \cdots U_1}); C^{cb}_S(\adm_{U_l \cdots U_1}) \le \delta \kappa(S) \big] \\
& \le 2\kappa(S) \mb P(C^{cb}_S(\adm_{U_l \cdots U_1}) \ge \delta \kappa(S)) + \delta \kappa(S),
\end{align*}
where in the last inequality we used the universal upper bound of $C_S^{cb}$ from Lemma \ref{lemma:elementary property} \eqref{universal upper bound}. Rearranging terms yields \eqref{ineqn: large time positive probability}.

Next, we set $\varepsilon = \delta = \frac{1}{4}$ and define $L = L_0\left(\frac{1}{4}\right)$. For any $l \le L$, there exists an integer $k$ such that $kl \ge L$ and $(k-1)l \le L$. This implies bounds on $\frac{1}{k}$:
\begin{align}\label{ineqn: bound on 1/k}
    \frac{l}{l+L} \le \frac{1}{k} \le \frac{l}{L}.
\end{align}
Using these bounds, we can write:
\begin{equation}\label{ineqn: replace k}
\mb P\big( C^{cb}_S(\adm_{U_l \cdots U_1}) \ge \frac{\kappa(S)}{4k} \big) \le \mb P\big( C^{cb}_S(\adm_{U_l \cdots U_1}) \ge \frac{l\kappa(S)}{4(l+L)} \big).
\end{equation}
Using the subadditivity of $C_S^{cb}$ under concatenation, we get:
\begin{equation}\label{ineqn: union bound}
\begin{aligned}
    \mb P\bigg( \bigcup_{i=1}^k \big\{C^{cb}_S(\adm_{U_{(i-1)l + 1} \cdots U_{(i-1)l + l}}) \ge \frac{\kappa(S)}{4k} \big\} \bigg) &\ge \mb P\big(C^{cb}_S(\adm_{U_{kl} \cdots U_1}) \ge \frac{\kappa(S)}{4} \big) \ge \frac{1}{4}.
\end{aligned}
\end{equation}
On the other hand, by the i.i.d. property of $\{U_l\}_{l\ge 1}$, for any $1 \le i \le k$, we have:
\begin{equation}\label{eqn: i.i.d}
\mb P\big( C^{cb}_S(\adm_{U_{(i-1)l + 1} \cdots U_{(i-1)l + l}}) \ge \frac{\kappa(S)}{4k} \big) = \mb P\big( C^{cb}_S(\adm_{U_l \cdots U_1}) \ge \frac{\kappa(S)}{4k} \big).
\end{equation}
Combining these results, we find:
\begin{align*}
    \frac{1}{4} & \underset{\eqref{ineqn: union bound}}{\le} \mb P\bigg( \bigcup_{i=1}^k \big\{C^{cb}_S(\adm_{U_{(i-1)l + 1} \cdots U_{(i-1)l + l}}) \ge \frac{\kappa(S)}{4k} \big\} \bigg) \\
    &\le \sum_{i=1}^k \mb P\big( C^{cb}_S(\adm_{U_{(i-1)l + 1} \cdots U_{(i-1)l + l}}) \ge \frac{\kappa(S)}{4k} \big) \\
    &\underset{\eqref{eqn: i.i.d}}{=} k \mb P\big( C^{cb}_S(\adm_{U_l \cdots U_1}) \ge \frac{\kappa(S)}{4k} \big) \\
    &\underset{\eqref{ineqn: replace k}}{\le} k \mb P\big( C^{cb}_S(\adm_{U_l \cdots U_1}) \ge \frac{l\kappa(S)}{4(l+L)} \big) \le k \mb P\big( C^{cb}_S(\adm_{U_l \cdots U_1}) \ge \frac{l\kappa(S)}{8L} \big).
\end{align*}
Finally, using \eqref{ineqn: bound on 1/k}, we have:
\begin{align*}
    \mb P\big( C^{cb}_S(\adm_{U_l \cdots U_1}) \ge \frac{l\kappa(S)}{8L} \big) \ge \frac{1}{4k} \ge \frac{l}{4(l+L)},
\end{align*}
which concludes the proof.
\end{proof}

\section{Proof of Theorem \ref{main:lower bound Hamiltonian simulation}}\label{app:proof of Hamiltonian simulation}

To prove the theorem, the first step is to adapt Theorem \ref{main:Gate count} to derive the following:

\begin{lemma}\label{lemma:intermediate Hamiltonian simulation}
For any $\delta \ge 0$,
\begin{equation}
    C_S^{cb}(\adm_{\exp(itH)}) \le \tau \cdot \overline{l}^S_{\delta}(\adm_{\exp(itH)}) + \delta \kappa(S),
\end{equation}
where $\kappa(S)$ is the quantum expected length defined in \eqref{def:expected length}. 
\end{lemma}

\begin{proof}
    The proof follows exactly the same as in Appendix \ref{app:proof of Gate count}, except for $\|\Psi_l^*(x) - x\|_{op} \le \tau l |||x|||_S, \forall x$, where 
    \begin{align*}
        \Psi_l = \int_{\mc U_{\tau}^l} \adm_{u_1}\cdots \adm_{u_l} d\mu^l(u_1,\cdots, u_l)
    \end{align*}
    minimizes $\overline{l}^S_{\delta}(\adm_{\exp(itH)})$. The additional factor $\tau$ arises because $C_S^{cb}(\adm_{\exp(ir H_{j})}) \le |r| \le \tau$.
\end{proof}

Now we are ready to prove the main theorem:

\begin{proof}[Proof of Theorem \ref{main:lower bound Hamiltonian simulation}]
Our goal is to show  
\begin{align*}
    C_S^{cb}(\adm_{\exp(itH)})  \ge \lambda_H t, \quad t \le \frac{1}{3\|H\|_{op}}.
\end{align*}
It suffices to present the lower bound for $C_S(\adm_{\exp(itH)})$. Using the definition, for any $x \notin S'$,
\begin{align*}
    & C_S(\adm_{\exp(itH)}) |||x|||_S \ge \|x - \adm_{\exp(itH)}^*(x)\|_{op} = \|x - e^{-iHt}xe^{iHt}\|_{op}.
\end{align*}
We aim to establish a lower bound on $\|x - e^{-iHt}xe^{iHt}\|_{op}$. Denoting $\delta_H^0(x) = x, \delta_H^{n+1}(x) = [H,\delta_H^n(x)]$, we have:
\begin{align*}
\|x - e^{-iHt} x e^{iHt}\|_{op} &= \|x - e^{it[H,x]}\|_{op} = \|i[H,x]t + \sum_{n\ge 2} \frac{(it)^n}{n!} \delta_H^n(x)\|_{op} \\
&\ge \|[H,x]\|_{op}t - \|\sum_{n\ge 2} \frac{(it)^n}{n!} \delta_H^n(x)\|_{op}.
\end{align*}
Moreover, 
\begin{align*}
\|\sum_{n\ge 2} \frac{(it)^n}{n!} \delta_H^n(x)\|_{op} &\le \sum_{n\ge 2} \frac{t^n}{n!} \|\delta_H^n(x)\|_{op} \le \sum_{n\ge 2} \frac{t^n (2\|H\|_{op})^{n-1}}{n!} \|[H,x]\|_{op} \\
&= (e^{2\|H\|_{op}t} - 1 - 2\|H\|_{op}t) \frac{\|[H,x]\|_{op}}{2\|H\|_{op}} \\
&\le (e-2)(2\|H\|_{op}t)^2 \frac{\|[H,x]\|_{op}}{2\|H\|_{op}} = 2(e-2)\|H\|_{op}\|[H,x]\|_{op}t^2,
\end{align*}
if $2\|H\|_{op}t \le 1$. Here, we used $\|\delta_H^n(x)\|_{op} \le (2\|H\|_{op})^{n-1} \|[H,x]\|_{op}$ in the second inequality and the estimate $e^a - a - 1 \le (e-2)a^2$ for $a \in [0,1]$ in the last inequality. Combining these results:
\begin{align*}
\|x - e^{-iHt}xe^{iHt}\|_{op} &\ge \|[H,x]\|_{op} t \big(1 - 2(e-2)\|H\|_{op} t\big) \ge \frac{\|[H,x]\|_{op}}{2}t, \quad \text{if } 4(e-2)\|H\|_{op}t \le 1.
\end{align*}
In summary, we showed:
\begin{equation}
C_S(\adm_{\exp(itH)}) |||x|||_S \ge \frac{\|[H,x]\|_{op}}{2}t,
\end{equation}
if $4(e-2)\|H\|_{op}t \le 1$. 

Finally, applying Lemma \ref{lemma:intermediate Hamiltonian simulation}, we have:
\begin{align*}
     \overline{l}^{\mc U_\tau}_{\delta}(\exp(itH)) &\ge \frac{C_S(\adm_{\exp(itH)}) - \delta \kappa(S)}{\tau} \ge \frac{\lambda_H t - \delta \kappa(S)}{\tau}.
\end{align*}
\end{proof}

\end{document}